\documentclass[12pt]{iopart}
\usepackage{graphicx}

\begin{document}

\title{Characteristic molecular properties of one-electron double quantum rings
under magnetic fields}

\author{JI Climente$^{1,2}$ and J Planelles$^2$}
\address{$^1$ CNR-INFM - National Research Center on nano-Structures and bio-Systems at Surfaces ($S3$), 
Via Campi 213/A, 41100 Modena, Italy}
\address{$^2$ Departament de Qu\'{\i}mica F\'{\i}sica i Anal\'{\i}tica, Universitat Jaume I, Box 224, E-12080 Castell\'o, Spain}

\ead{josep.planelles@qfa.uji.es}

\date{\today}

\begin{abstract}

The molecular states of conduction electrons in laterally coupled quantum 
rings are investigated theoretically.
The states are shown to have a distinct magnetic field dependence, which
gives rise to periodic fluctuations of the tunnel splitting and ring
angular momentum in the vicinity of the ground state crossings.
The origin of these effects can be traced back to the Aharonov-Bohm 
oscillations of the energy levels, along with the quantum mechanical 
tunneling between the rings. 
We propose a setup using double quantum rings which shows that 
Aharonov-Bohm effects can be observed even if the net magnetic flux 
trapped by the carriers is zero.

\end{abstract}


\maketitle

%%%%%%%%%%%%%%%%%%%  Intro  %%%%%%%%%%%%%%%%%%%%%%%%%%%%%%%
\section{Introduction}

When the energy levels of two tunnel-coupled semiconductor quantum dots 
are set into resonance, the carriers localized in the individual
nanostructures hybridize forming molecular-like states, in good analogy
with atomic molecular bonds\cite{KouwenhovenSCI,HolleitnerSCI,
AlivisatosNAT,BlickPRL,AustingPB,BayerSCI}.
The possibility to engineer the properties of these 'artificial molecules',
such as the bond length, the material and shape of the constituent 'atoms' 
or the number and nature of delocalized carriers has opened new paths 
to learn basic physics of molecular systems\cite{PiPRL,DotyXXX,DybalskiPRB}.
 
Vertically-coupled\cite{AustingPB,BayerSCI,PiPRL,DotyXXX} and 
laterally-coupled\cite{ChaJKP,StanoPRB,AbolfathJCP,vanderWielNJP,
PettaPE,KoppensNAT,vanderWielRMP} double quantum dots (DQDs) 
have received particular attention owing to additional technological 
implications for the development of scalable two-qubit logic gates\cite{LossPRA}.
Much less is known about other kinds of artificial molecules, such as 
double quantum rings (DQRs). This is nonetheless an interesting problem,
as the remarkable magnetic properties of semiconductor quantum rings (QRs)\cite{LeePE},
related to the Aharonov-Bohm (AB) effect\cite{AharonovPR}, have been 
thoroughly studied at a single-ring level\cite{ViefersPE}, but their effects 
at a molecular level remain largely unexplored.

Molecular states in vertically coupled QRs have been investigated\cite{SuarezNT,
GranadosAPL,ClimentePRB,LiJJAP,MaletPRB,CastelanoPRB,PiacenteJAP}, 
but the hybridized orbitals in such systems are aligned along the vertical direction. 
As a result, their response to vertical magnetic fields, which are responsible for
AB effects, is very weak\cite{ClimentePRB}.
The electronic states of concentrically coupled QRs have also
been studied\cite{ManoNL,FusterBJP,PlanellesEPJB,SzafranPRB,ClimentePRB2,MaletPRB2}.
However, the markedly different vertical confinement of the inner
and outer rings leads to carrier localization in individual rings, preventing molecular 
hybridization\cite{ManoNL,FusterBJP,PlanellesEPJB,MaletPRB2}.

Recently, we studied the molecular dissociation of yet another kind of structure, 
namely laterally-coupled QRs.\cite{PlanellesJPCS}
Interestingly, in such structures the two rings may have similar dimensions, 
which grants the formation of covalent molecular orbitals.
In addition, tunnel-coupling takes place in the plane of the rings, thus 
rendering molecular orbitals sensitive to vertical magnetic fields. 
These two ingredients make laterally-coupled QRs ideal systems to attain 
magnetic modulation of molecular bonds and their derived properties.
This is the subject of research in the present paper.
We investigate the energy structure of DQRs under magnetic fields, and
find that the AB-induced ground state crossings lead to sudden maxima of 
the tunnel splitting between bonding and antibonding orbitals of DQRs, as well
as to periodic suppressions of the carrier rotation within the rings.

Laterally-coupled DQRs also offer the possibility to extend the research
of AB effects, typically restricted to effectively isolated ring structures
(see e.g. Refs.~\cite{BayerPRL,LorkePRL,FuhrerNAT,ChandrasekharPRL}),
to composite systems, which may unveil subtleties of these purely quantum 
mechanical phenomena. As a matter of fact, here we show that, unlike in 
single QRs\cite{AharonovPR,ViefersPE}, AB effects in composite systems need 
no finite net magnetic flux to take place.

The paper is organized as follows. In Section \ref{s:theo} we describe the
theoretical model employed and give details of the DQR structure under 
consideration.  In Section \ref{s:res} results are reported: in Section 
\ref{ss:tun}, the energy structure and tunnel splitting of DQRs are studied;
in Section \ref{ss:cur}, the expectation value of the rings angular momentum
is used as an estimate of the carrier rotation within the rings; finally,
in Section \ref{ss:AB} the AB oscillations of electrons in single and DQRs 
are compared. Conclusions are presented in Section \ref{s:conc}.

\begin{figure}[h]
\begin{center}
\includegraphics[width=0.75\textwidth]{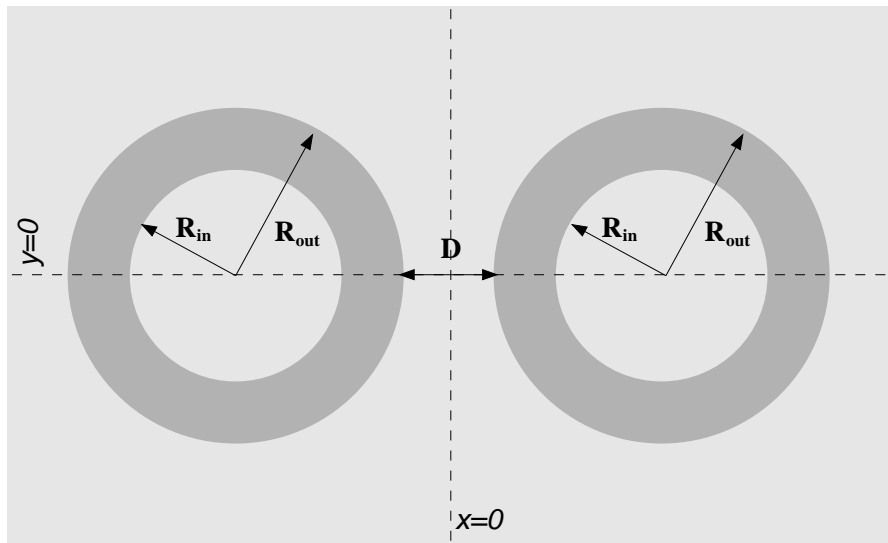}
\caption{Schematic of the DQR structure under study and the 
relevant geometrical parameters. The confining potential is
zero inside the rings and $V_c$ outside.}\label{fig1}
\end{center}
\end{figure}

%%%%%%%%%%%%%%%%%%% Theory  %%%%%%%%%%%%%%%%%%%%%%%%%%%%%%%
\section{Physical system and theoretical model}
\label{s:theo}

We consider QRs in the Coulomb-blockade regime, charged with only one
conduction electron. QRs in this regime may be fabricated using
either self-assembly\cite{LeePE} or litographic\cite{BayerPRL}
techniques.
However, control of the nanostructure dimensions and positioning 
is only accurate when using litographic methods. Therefore, for illustration 
purposes in this work we choose to simulate a DQR system built from etched QRs, 
as those of Ref.~\cite{BayerPRL}. 

Since QRs have much stronger vertical than lateral confinement,
we calculate the low-lying states of the DQRs using a two-dimensional
effective mass-envelope function approximation Hamiltonian which 
describes the in-plane ($x-y$) motion of the electron in the
ring. In atomic units, the Hamiltonian may be written as:

\begin{equation}
\label{eq1}
H=\frac{1}{2 m^*} (\mathbf{p}+\mathbf{A})^2+V(x,y),
\end{equation}

\noindent where $m^*$ stands for the electron effective mass, $\mathbf{p}$
is the canonical moment, and $V(x,y)$ is a square-well potential
confining the electron within the DQR structure shown in Fig.~\ref{fig1}. 
In polar coordinates it has the compact expression 
$V(\rho,\theta)=0$ if $R_{in} < \rho < R_{out}$ and
$V(\rho,\theta)=V_c$ elsewhere, with $V_c$ as the barrier 
confining potential. 
$\mathbf{A}$ is the vector potential.  
Unless otherwise stated, we use the symmetric gauge, $A=B/2\,(-y,x,0)$, 
which introduces a magnetic field $B$ pointing along the growth direction $z$. 
Replacing this vector potential into the Hamiltonian, one obtains:

\begin{equation}
\label{eq2}
H=\frac{\hat p_{\parallel}^2}{2 m^*} + \frac{B^2}{8 \; m^*} (x^2+y^2)
- i \frac{B}{2 \; m^*} (x \frac{\partial}{\partial y}-y \frac{\partial}{\partial x}) + V(x,y),
\end{equation}

\noindent where ${\hat p_{\parallel}}^2={\hat p_x}^2+{\hat p_y}^2$.
The eigenvalue equation of Hamiltonian (\ref{eq2}) is solved numerically using
a finite-difference scheme on a two-dimensional grid ($x,y$) extended far beyond 
the DQR limits. 

Following Ref.~\cite{BayerPRL}, the QRs we study are made of In$_{0.1}$Ga$_{0.9}$As 
and they are surrounded by GaAs barriers.  Reasonable material parameters for this 
heterostructure are barrier confinement potential $V_c=50$ meV\cite{BayerPRL}
and effective mass $m^*=0.05$\cite{KorkusinskiPSSb}.
% BayerPRL says Vc = few tens of meV
Unless otherwise stated, the inner radius of the rings is $R_{in}=15$ nm, 
the outer one $R_{out}=45$ nm, and the two rings are separated horizontally 
by a $D=3$ nm barrier.

%%%%%%%%%%%%%%%%%%% Results  %%%%%%%%%%%%%%%%%%%%%%%%%%%%%%%
\section{Results}
\label{s:res}

\subsection{Energy structure and tunnel splitting}
\label{ss:tun}

\begin{figure}[h]
\begin{center}
\includegraphics[width=0.75\textwidth]{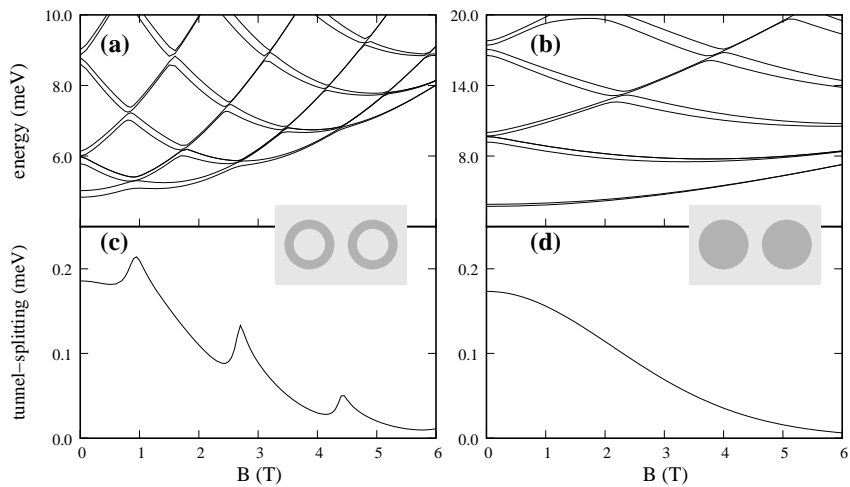}
\caption{Top row: energy levels vs magnetic field for a DQR (a) and a DQD (b).
Bottom row: tunnel splitting vs magnetic field for a DQR (c) and a DQD (d).
The instets provide schematic representations of the structures under study.
Note the non-monotonic evolution of the tunnel splitting for DQRs.}\label{fig2}
\end{center}
\end{figure}

In this section we analyze the energy structure and tunnel-coupling of a
DQR under magnetic fields, and compare them to the well-known case of DQDs. 
The DQD has interdot barrier $D=3$ nm and $R_{out}=30$ nm, which gives
a similar area to that of the DQR.
The energy structure of both kinds of artificial molecule are illustrated 
in the top panels of Fig.~\ref{fig2}.
As is known, the DQD energy structure resembles the spectrum of a
quantum disk, but with two-fold levels corresponding to the
bonding and antibonding linear combinations of the single dot 
states\cite{ChaJKP,StanoPRB}.
In a similar fashion, the DQR energy structure resembles that of a 
single QR, with the usual AB oscillations, but the levels are two-fold
as corresponding to the bonding and antibonding molecular 
states\cite{PlanellesJPCS}.

The energy spacing between bonding and antibonding energy levels, 
the so-called tunnel splitting, indicates the strength of the molecular bonds,
and it has important practical implications for DQD devices, where
it determines the degree of quantum entanglement\cite{BayerSCI}
and affects electron transport\cite{HuttelPRB}.
In the bottom row of Fig.~\ref{fig2} we compare the ground state
tunnel splitting of the DQD and DQR.

For the DQD, the tunnel splitting decreases monotonically with 
increasing field, in agreement with experimental data\cite{HuttelPE} 
and previous theoretical works\cite{ChaJKP}. 
This is because a higher field implies smaller Landau orbits, i.e.
the field squeezes the electron wavefunction within the dots, 
thus reducing the amount of charge in the inter-ring barrier.
For DQRs, however, the tunnel splitting no longer exhibits
a monotonic behaviour. In general, it also decreases with growing
field, due to wavefunction squeezing, but in addition it shows 
sudden peaks at quasi-periodic values of $B$. 
Clearly, the values of the field where such peaks occur correspond to the 
level crossings in the energy structure (Fig.~\ref{fig1}(a)), i.e. to 
integer number of AB periods\cite{AharonovPR}.
This tunneling enhancement, whose origin we explain below, suggests
that DQRs enable a stronger magnetic field-induced modulation of the molecular
strength than DQDs, and larger tunnel splittings may be achieved when
operating at finite magnetic fields, which may be of interest for
spin qubit systems, where magnetic fields are used to manipulate spin 
states\cite{vanderWielNJP,PettaPE,KoppensNAT,LossPRA}.
We point out that this behaviour is characteristic for laterally coupled DQRs. 
In vertically coupled QRs, tunnel splitting is constant against
the field\cite{ClimentePRB,LiJJAP}.

\begin{figure}[h]
\begin{center}
\includegraphics[width=0.75\textwidth]{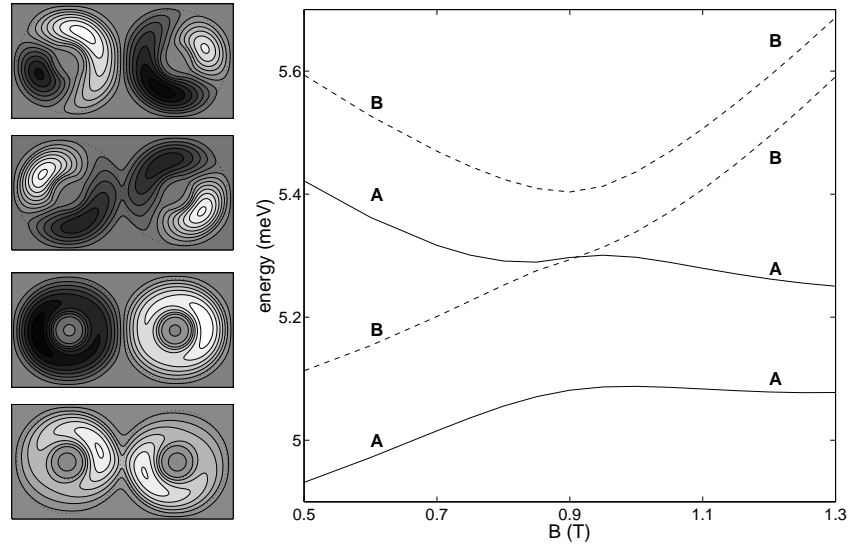}
\caption{Right side: zoom of the four lowest-energy levels of Fig.~\ref{fig2} 
DQR in the vicinity of the first anticrossing. 
Solid (dashed) lines are used for states with irreducible representation A (B). 
Point group symmetry is $C_2$.
Left side: contours of the wavefunction real part for the first 
(bottom panel) to the fourth (top panel) energy level at $B=0.5$ T.
White stands for positive and black for negative wavefunction.
Dotted lines indicate the DQR limits.}\label{fig3}
\end{center}
\end{figure}

To understand the large values of the DQR tunnel splitting in the vicinity 
of level crossings, in Fig.~\ref{fig3} (right panel) we zoom in on the 
lowest-energy levels around the first crossing point. 
At this point, in a single QR one would expect a series of level crossings\cite{ViefersPE}. 
However, Fig.~\ref{fig3} reveals a crossing between the second and third 
levels only, the first and fourth states being pushed away by apparent
anticrossings.
This is because the point group symmetry of a QR under vertical magnetic
fields, $C_{\infty}$, is lowered to $C_2$ in a DQR.
The DQR electron states are then classified by the irreducible 
representations A and B, indicating even and odd symmetry under a
rotation of $\pi$ degrees, respectively.
The symmetry of each level can be ellucidated by inspecting the 
wavefunctions before the anticrossing (left side of Fig.~\ref{fig3}).
The energy increases from bottom (ground state) to top 
(fourth state). Only the real part of the wavefunction is 
illustrated, as it suffices to capture the relevant aspects
of the symmetry. 
It is clear that the first and second (third and fourth) levels 
form bonding and antibonding molecular states built from the same 
'atomic' orbitals.
On the other hand, it can be seen that the first and third levels 
have symmetry A (even) while the second and fourth ones have symmetry B (odd).
Therefore, the symmetry sequence of the four lowest-lying levels 
shown in Fig.~\ref{fig3} is A B A B before the anticrossing,
and A A B B after it.

Since the second and third states have different symmetry,
they simply cross each other. By contrast, the first and third
(second and fourth) states have the same symmetry, so that 
they undergo anticrossings.
The anticrossings prevent a smooth magnetic field dependece
of the tunnel splitting.
Hence the non-monotonic evolution of Fig.~\ref{fig2}(b).

\begin{figure}[h]
\begin{center}
\includegraphics[width=0.75\textwidth]{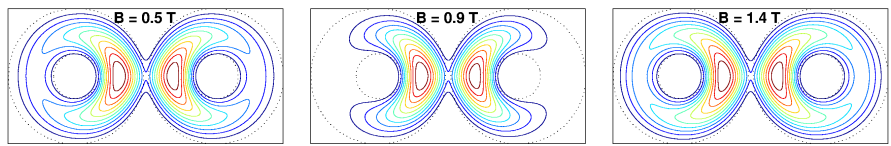}
\caption{(Colour online). Contour of the ground state charge 
density before (left panel), during (central panel) and after (right panel)
the anticrossing. 
Dotted lines indicate the DQR limits.
Note the enhanced tunnel-coupling of the ground state at the
anticrossing point ($B=0.9$ T).}\label{fig4}
\end{center}
\end{figure}

Insight into the tunnel-coupling strenght can be obtained by
observing Fig.~\ref{fig4}, where we plot the charge density
of the ground state before, during and after the anticrossing.
At the anticrossing point, where the interaction between
the first and third levels is at its maximum, the charge is pushed
towards the inter-ring barrier, leading to the tunnel-coupling 
enhancement reported in Fig.~\ref{fig2}(b).
 
\subsection{Expectation value of angular momentum}
\label{ss:cur}

\begin{figure}[h]
\begin{center}
\includegraphics[width=0.75\textwidth]{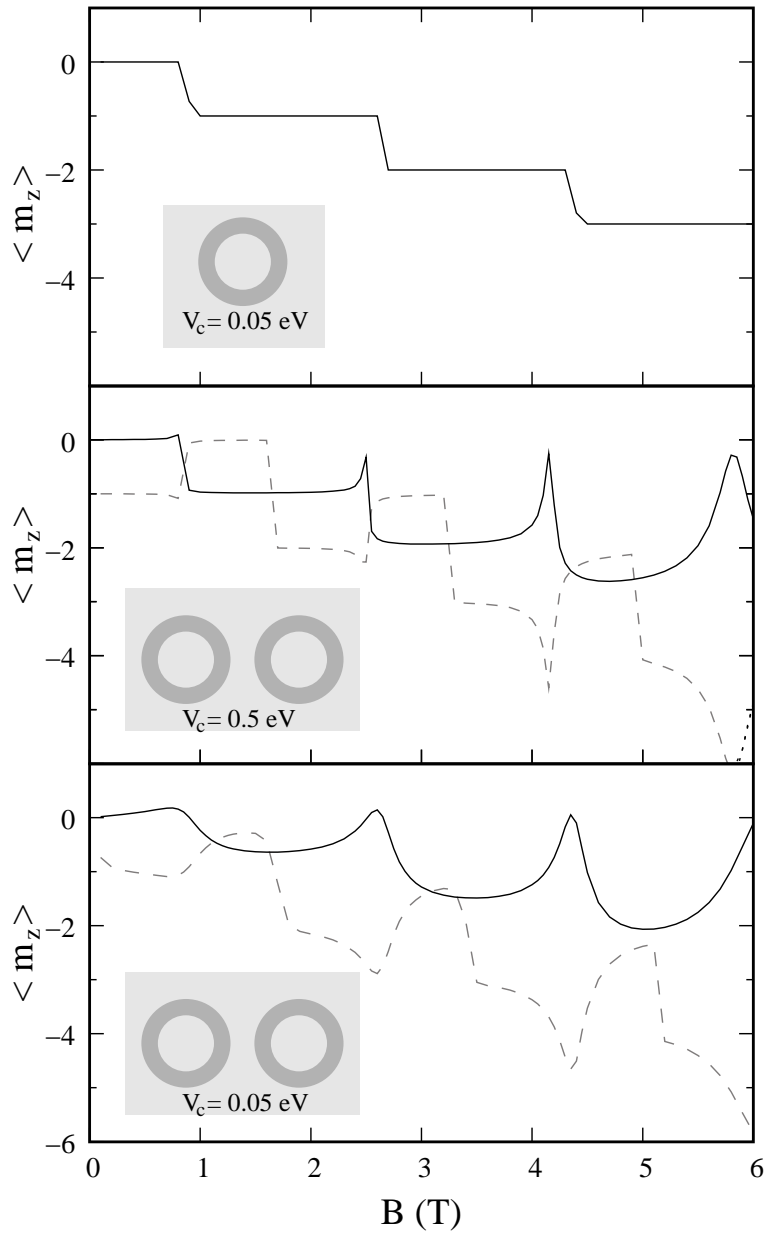}
\caption{Expectation value of the ground state angular momentum 
vs.~magnetic field in (a) a single QR, (b) a DQR with high inter-ring 
barrier, and (c) a DQR with regular inter-ring barrier height. 
The dashed lines in (b) and (c) represent the expectation value of 
the third level. The insets are schematics of the structures.}\label{fig5}
\end{center}
\end{figure}

An intrinsic property of QRs, closely related to the AB effect, is the 
appearance of one-electron ground states with finite angular momenta, which
give rise to an equilibrium current arising from carrier rotation within
the structure\cite{ViefersPE}.
In isolated QRs, this current is proportional to the ground state azimuthal
angular momentum $m_z$.
In DQRs, due to the lowered symmetry, the angular momentum is no longer a 
good quantum number.  However, the expectation value of the angular momenta
within each of the constituent rings can still be taken as a 
measurement of the carrier rotation within the nanostructure.
Thus, in Fig.~\ref{fig5} we compare $\langle m_z \rangle$ for a single QR, 
a DQR with high ($V_c=500$ meV) inter-ring barrier and a DQR with regular 
(experimental-like) barrier height.\footnote{For the DQR structures, 
the $\langle m_z \rangle$ is calculated as the sum of the left and right 
QR local angular momenta, i.e. the expectation values defined from the 
origin of each QR.}

In the single QR case, Fig.~\ref{fig5}(a), $\langle m_z \rangle$ is a 
step function, decreasing in one unit of angular momentum with every 
ground state crossing\cite{ViefersPE}.
A similar behaviour is found in the DQR case with high inter-ring
barrier, solid line in Fig.~\ref{fig5}(b), as the QRs are almost isolated.
However, a new feature appears in the vicinity of the level crossings
($B=2.5, 4.1$ and $5.7$ T). Here, $\langle m_z \rangle$ rapidly goes 
to zero, which implies as a sudden quenching of the carrier rotation.
The introduction of full tunnel-coupling in the DQR, solid line in
Fig.~\ref{fig5}(c), further modifies the magnetic field dependence
of $\langle m_z \rangle$.
First, $\langle m_z \rangle$ no longer takes integer values.
Instead, it takes fractional values, reduced as compared to the case of 
weakly-coupled rings.
This is because the charge density in the tunnel barrier, which is
now increased, does not contribute to the rotation within the rings.
Second, $\langle m_z \rangle$ goes to zero every time an AB 
period is completed, in such a way that the periodic quenching of 
the ring current starts from weak magnetic fields.
 
The origin of the $\langle m_z \rangle$ peaks in Fig.~\ref{fig5}(b) and (c)
is the interaction between the first and third levels of the DQR at the 
anticrossing points, discussed above.
Away from the anticrossing, the first and third levels have angular momenta 
$\langle m_z \rangle_1$ and $\langle m_z \rangle_3$, solid and dashed lines in 
Fig.~\ref{fig5}.
Since these levels are the two lowest bonding orbitals, their angular momenta
are similar to those of the two lowest levels in a single QR.
%This is particularly clear in Fig.~\ref{fig5}(b), where the charge density
%in the barrier is small, so that $\langle m_z \rangle_1$ evolves with the field 
%as $\langle m_z \rangle_1 \approx 0,-1,-2\ldots$.
%
At the anticrossing, however, the strong interaction couples 
$\langle m_z \rangle_1$ and $\langle m_z \rangle_3$ in such a way that the 
ground state tends to 
$\langle m_z \rangle_{gs}=\langle m_z \rangle_1 - \langle m_z \rangle_3$
and the excited state to
$\langle m_z \rangle_{ex}=\langle m_z \rangle_1 + \langle m_z \rangle_3$.
Thus, at the first anticrossing $\langle m_z \rangle_{ex} \approx 0+(-1) = -1$,
at the second one $\langle m_z \rangle_{ex} \approx -1+(-2) = -3$,
at the third one $\langle m_z \rangle_{ex} \approx -2+(-3) = -5$, etc.
By contrast, the ground state does not reach the expected $\langle m_z \rangle_1 
- \langle m_z \rangle_3$ value, because it tends to deposit a large amount of 
charge density in the tunneling barrier (recall  Fig.~\ref{fig4} center). 
This provides maximum stabilization for the (bonding) ground state, 
but it also leads to $\langle m_z \rangle_{gs} \approx 0$.\\
%Note that, in the vicinity of the anticrossings, the total angular momentum 
%expectation of the subsystem formed by the first and third levels is 
%approximately conserved, as revealed by the complementary shape of 
%the solid and dashed curves.\\

In typical QR structures, the physical observable is the persistent 
current\cite{KleemansPRL}, which is proportional to the 
magnetization\cite{CheungPRB}:

\begin{equation}
M=-\frac{\partial E_{gs}}{\partial B}.
\end{equation}

\begin{figure}[h]
\begin{center}
\includegraphics[width=0.75\textwidth]{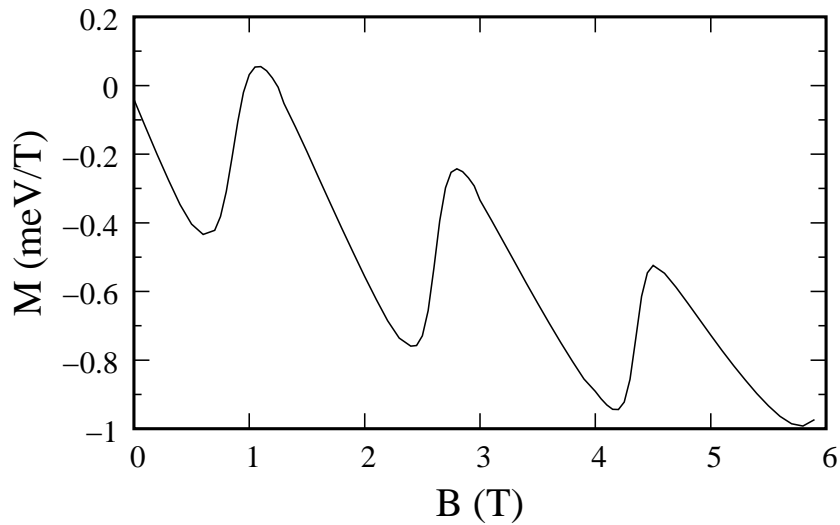}
\caption{Magnetization of DQR. The $C_2$ symmetry is responsible for
the rounded edges of the curve.}\label{fig6}
\end{center}
\end{figure}

\noindent Here $E_{gs}$ is the ground state energy.
The persistent current includes not only the current arising from the 
electron angular momentum, but also that coming from the interaction with 
the magnetic field vector potential\cite{1D}.
Thus, rather than presenting complete suppressions, the persistent
current looks as in Fig.~\ref{fig6}, where the magnetization is represented.
While in single QRs the magnetization would give a saw-tooth 
picture\cite{CheungPRB,ChakraPRB,ClimentePRB3}, here rounded edges
are obtained. This is a usual signature of lowered symmetry, which
allows additional interactions among the states\cite{PlanellesNT}.
The interactions are particularly strong at the crossing points,
thus producing the rounded edges.

\subsection{AB oscillations in DQRs with zero net magnetic flux}
\label{ss:AB}

The energy level oscillations of charged carriers confined in QRs under
magnetic fields, as those shown in Fig.~\ref{fig2}(a), constitute 
a manifestation of the AB effect, i.e. an action upon the quantum system 
exerted by the vector potential $\mathbf{A}$\cite{AharonovPR}.
The action is induced by the magnetic flux enclosed in the
trajectory described by the carrier, $\Phi=\oint {\mathbf A}\,d{\mathbf l}=
\int {\mathbf B}\,d{\mathbf S}$.
Every time this flux is a multiple of the unit flux quantum, 
$\Phi_0=2 \pi \hbar / e$, the energy spectrum retrieves the zero field
structure\cite{diamg} and the system is said to accumulate one AB phase 
unit\cite{ViefersPE}.
Experimental evidence of such AB oscillations have been reported
in different types of mesoscopic\cite{FuhrerNAT} and nanoscopic semiconductor 
QRs.\cite{BayerPRL,LorkePRL}.
 
In this section, we report on the (to our knowledge) first study of AB
effect in composite systems, and reveal a new aspect of this phenomenon,
namely that AB oscillations can be found even when the total flux trapped
by the carriers is $\Phi=0$, provided the flux threading the individual
rings is finite.
To this end, we consider two magnetic field configurations:
\begin{enumerate}
\item a uniform positive magnetic field goes through the 
entire system. Hereafter we refer to this as parallel field, $B_{p}$.
\item a local positive field goes through the right side of the
structure, while a negative one goes through the left one.
Hereafter we refer to this as antiparallel field, $B_{a}$.
\end{enumerate}
The two cases can be modeled using a Coulomb gauge vector 
potential ${\mathbf A}=B\,(0,x,0)$ for the right side of the plane,
and ${\mathbf A}=\pm B\,(0,x,0)$ for the left one, where the positive
(negative) sign applies for parallel (antiparallel) fields.
Note that this gauge choice grants continuity of ${\mathbf A}$ 
and $\nabla {\mathbf A}$ in the whole domain
The Hamiltonian now reads:

\begin{equation}
\label{eq3}
H=\frac{\hat p_{\parallel}^2}{2 m^*} 
+ \frac{B^2}{2 \; m^*} x^2
\pm i \frac{B}{m^*}\, x \frac{\partial}{\partial y} 
+ V(x,y),
\end{equation}

\noindent where the negative sign of the linear term in $B$
applies for $B_a$ and $x<0$.

\begin{figure}[h]
\begin{center}
\includegraphics[width=0.75\textwidth]{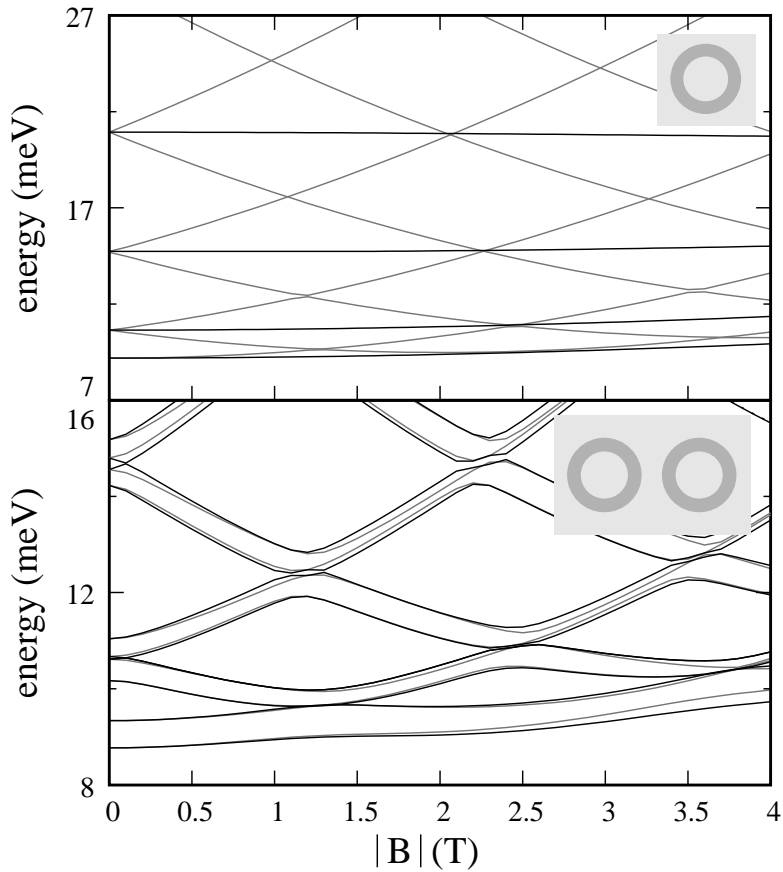}
\caption{Energy levels of a single QR (top panel) and a DQR (bottom panel) 
vs.~absolute value of the magnetic field. 
Black (gray) lines stand for the case of antiparallel (parallel)
fields applied on the left and right halves of the structure (see text). 
The insets are schematics of the structures under study.
Note that parallel and antiparallel fields yield dramatically 
different responses in a single QR, but almost identical in a DQR.}\label{fig7}
\end{center}
\end{figure}

We diagonalize Eq.~(\ref{eq3}) to calculate the energy structure of a 
single QR and that of a DQR under parallel and antiparallel magnetic fields. 
The results, corresponding to rings with $R_{in}=15$ nm and $R_{out}=35$ nm 
($D=3$ nm for the DQR), are shown in Fig.~\ref{fig7}.
For a single QR, while $B_p$ (gray lines) yields usual 
AB oscillations, $B_a$ (black lines) yields a featureless 
spectrum, affected by the diamagnetic shift only.
This is the expected difference between the cases where finite and null
magnetic flux is trapped by the carrier.
A strikingly different response is however obtained for the DQR,
as the energy structure looks almost the same regardless of the 
magnetic field configuration.
This is surprising, because in the antiparallel case, the magnetic 
flux penetrating the left and right rings cancels out, so that the net 
flux enclosed by the electron is again zero, and one may not expect 
AB manifestations.

\begin{figure}[h]
\begin{center}
\includegraphics[width=0.7\textwidth]{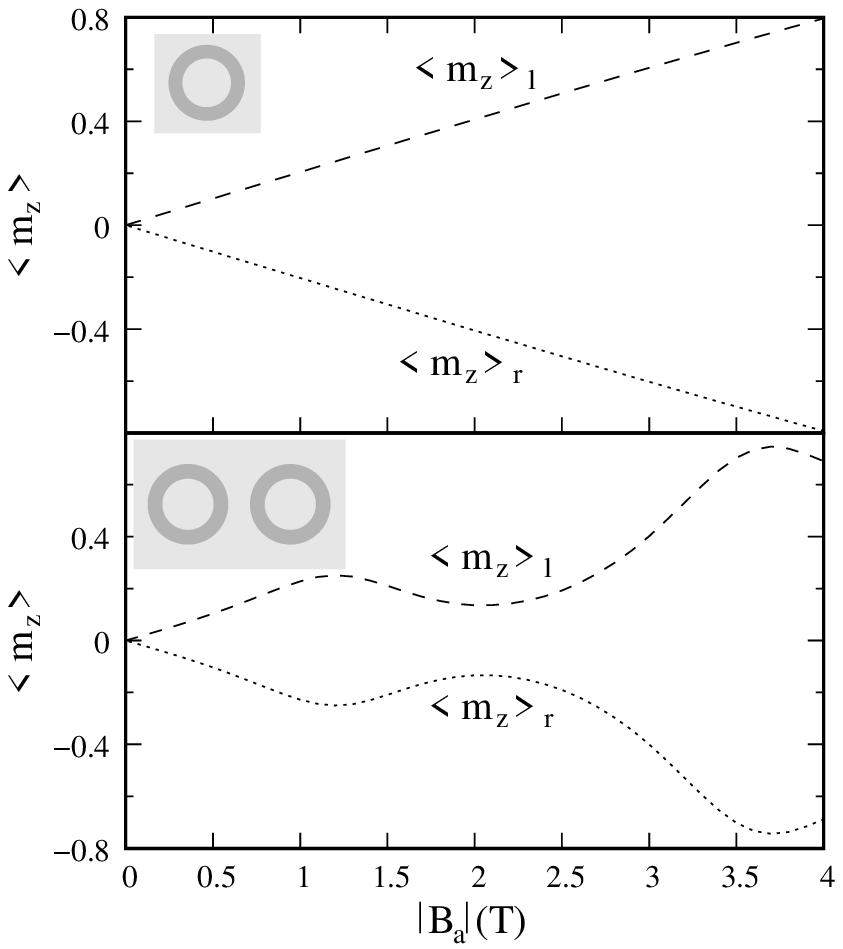}
\caption{Ground state angular momentum expectation value of the left and right 
halves of the structure vs absolute value of the antiparallel magnetic field.
Top row: single QR. Bottom row: DQR. 
Dashed and dotted lines are used for $\langle m_z \rangle_l$ 
and $\langle m_z \rangle_r$, respectively.
The insets are schematics of the structures.}\label{fig8}
\end{center}
\end{figure}

To gain some insight into the different behavior of the single and
double QR, in Fig.~\ref{fig8} we depict the angular momentum
expectation value for the left ($\langle m_z \rangle_l$) and right 
($\langle m_z \rangle_r$) halves of the each structure, as a function
of $B_a$.
Clearly, the antiparallel field induces opposite left and right 
angular momenta for both structures. Therefore, 
$\langle m_z \rangle = \langle m_z \rangle_l + \langle m_z \rangle_r=0$,
which is consistent with the systems picking a net zero AB phase\cite{ViefersPE}.
Yet, only the single QR energy spectrum shows no AB oscillations.

This can be interpreted as follows. The antiparallel field induces 
clockwise and anticlockwise carrier rotation in each half of the structure.
For a single QR, the two currents cancel each other out, the angular momentum
of the ring is always zero and then the states are insensitive to the linear 
term of the magnetic field (responsible for the AB effects).
By contrast, for a DQR, the are \emph{finite} (though opposite) currents in 
each of the rings. 
Thus, if the left and right rings were uncoupled, it is immediate that both 
would trap magnetic flux and hence show AB oscillations.
Moreover, from the $C_2$ symmetry of Hamiltonian (\ref{eq3}), 
it is easy to show that the energy spectrum of the two uncoupled QRs would 
be identical but with reversed sign of the angular momenta (in other words, 
the currents are identical in magnitude but opposite in sign).

\begin{figure}[h]
\begin{center}
\includegraphics[width=0.75\textwidth]{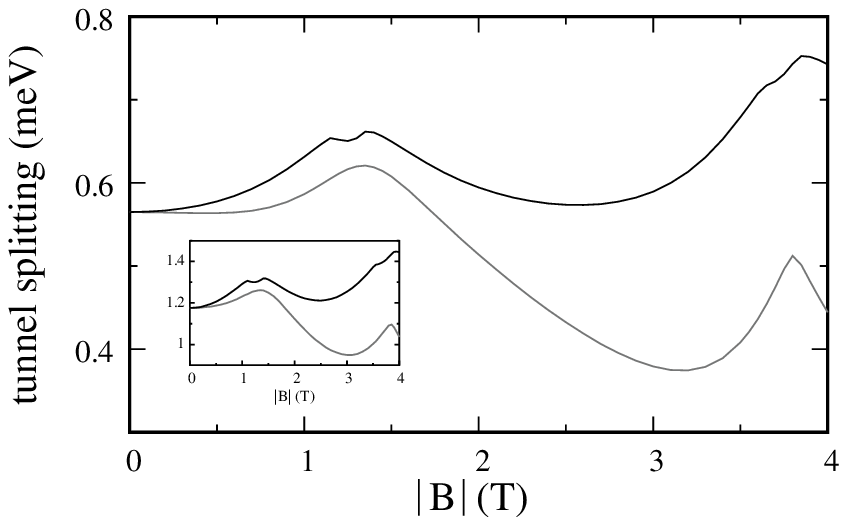}
\caption{Tunnel splitting vs absolute value of the magnetic field for a DQR with 
$D=3$ nm.  Black and gray lines are for antiparallel and parallel magnetic fields.
The inset shows the results for a DQR with $D=1$ nm (enhanced tunnel-coupling).}\label{fig9}
\end{center}
\end{figure}

Switching on the tunnel-coupling enables the electron to delocalize over
the two rings while keeping the net trapped magnetic flux zero.
Still, as can be seen in the bottom panel of Fig.~\ref{fig7}, the spectra 
under parallel or antiparallel fields remain similar.
%Only small quantitative deviations are found. 
This is because the tunnel-coupling for the two magnetic field configurations
is qualitatively similar (although not identical) as can be seen in Fig.~\ref{fig9},
where gray lines represent the tunnel splitting under $B_p$ and black
lines that under $B_a$.
Note also that, in the antiparallel field case, the tunnel splitting increases 
with $|B|$, this being responsible for the differences in the energy spectra of 
Fig.~\ref{fig7}.
The same behaviour is found in DQRs with stronger tunnel-coupling, as shown 
in Fig.~\ref{fig9} inset, where the tunnel splitting of DQRs with a thin
($D=1$ nm) barrier is plotted.

We then conclude that both the individual ring energies and the tunnel-coupling
are similar for $B_p$ and $B_a$, which explains the appearence of similar 
spectra in Fig.~\ref{fig7}.

\begin{figure}[h]
\begin{center}
\includegraphics[width=0.75\textwidth]{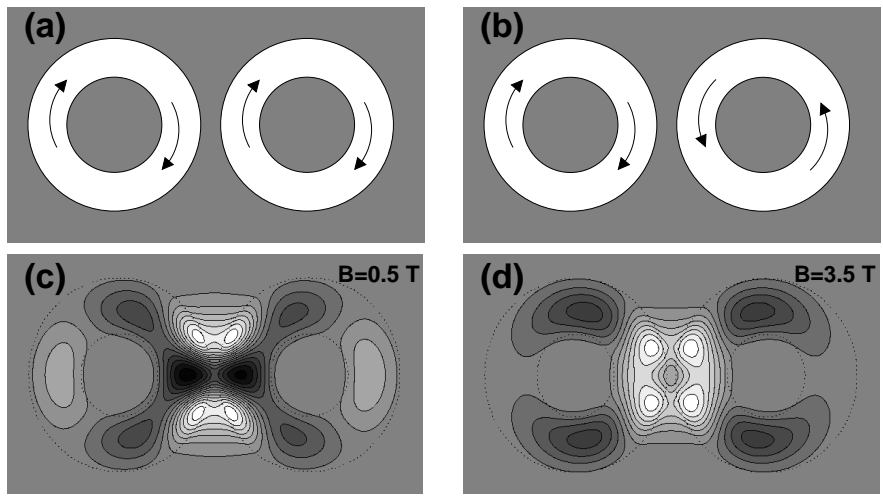}
\caption{
(a) and (b): schematics of the currents induced by parallel and
antiparallel fields, respectively.
(c) and (d): contour of the charge density difference between
the antiparallel field ground state and the parallel field one, 
under weak and strong magnetic fields, respectively. 
White (black) regions indicate excess of antiparallel (parallel) field 
state charge. Dotted lines show the DQR limits. Note that antiparallel 
states have larger charge density in the tunneling region for strong 
magnetic fields.}\label{fig10}
\end{center}
\end{figure}

Last, we comment on the fact that the tunnel splitting increases with 
$|B|$ in the DQR subject to antiparallel field, Fig.~\ref{fig9}. 
This is an anomalous behaviour, because the increasing magnetic confinement 
should lead to reduced charge density in the barrier and hence to decreasing
tunnel splitting, as in the parallel field case\cite{HuttelPE}. 
This can be explained in terms of the sense of the carrier circulation in each ring. 
For parallel field, the electron rotates in the same sense (say clockwise) 
in the two rings.
As a result, the current in the surroundings of the tunneling barrier
is different for each ring (see schematic representation in Fig.~\ref{fig10}(a)).
This hinders the charge sharing between the subsystems.
By contrast, for antiparallel field, the sense of rotation in the left 
and right rings is opposite. 
As a result, the current in the surroundings of the tunneling barrier 
is now the same (Fig.~\ref{fig10}(b)).
This in turn favours the charge density sharing.
With increasing field, the current grows and these trends become 
more important\cite{ow}. Indeed, for antiparallel field, the favoured 
charge sharing is able to compensate for the wavefunction squeezing.

To illustrate this effect, in Fig.~\ref{fig10}(c) and (d) we plot the 
difference in charge density between $B_a$ and $B_p$ ground states,
for weak ($B=0.5$ T) and strong ($B=3.5$ T) magnetic fields.
White contour regions indicate excess charge density coming from the
antiparallel case, and black regions vice-versa.
When the field is weak, the charge density in the barrier is similar
for the two systems. However, for the strong field, it is apparent that
the antiparallel ground state deposits much more density in the tunneling
barrier, which results in its enhanced tunnel-coupling.

\section{Conclusions}
\label{s:conc}

We have shown that the electron states in side-coupled coupled QRs
display a characteristic behaviour when subject to vertical magnetic
fields, different from that observed in other QR and QD structures. 
In particular, the tunnel-coupling strength is found to oscillate
with the field, showing sharp maxima when the AB effect induces
ground state crossings. This may be of interest e.g. for strong
magnetic modulation of the transport probability between the 
nanostructures.
In these tunnel-coupling maxima, the carrier rotation within the
rings is abruptly supressed, owing to charge accumulation in the
inter-ring barrier. This introduces a characteristic magnetic field
dependence of the persistent currents which may be verified 
experimentally.\cite{KleemansPRL}

We have also shown that DQRs, as quadruply-connected systems, 
may reveal new fundamental aspects of quantum physics arising from 
the AB effect. 
In the single QR structures (doubly-connected systems) investigated 
to date, a non-zero magnetic flux piercing the loop is required 
to produce AB effects, such as AB oscillations.  
From our theoretical prediction, it follows that this is not
necessarily the case for DQRs, where AB oscillations are present 
even if the net magnetic flux piercing the two loops (and 
hence the total accumulated AB phase) is zero,
provided the flux going through the individual rings is finite.
The experimental setup to prove this should consist of two tunnel-coupled
QRs subject to antiparallel magnetic fields in the left and right rings.
DQR structures can be currently realized with remarkable precision 
using litographic techniques, as in Refs.~\cite{BayerPRL,FuhrerNAT}, 
but the antiparallel field realization may be more challenging. 
Laser-controlled currents might provide a feasible alternative\cite{PershinPRB}.

\ack We acknowledge support from MEC-DGI projects CTQ2004-02315/BQU, 
UJI-Bancaixa project P1-1B2006-03 and Cineca Calcolo Parallelo 2007.
One of us (J.I.C.) has been supported by the EU under the Marie Curie 
IEF project MEIF-CT-2006-023797.


\begin{thebibliography}{50}
\bibitem{1D}
In strictly one-dimensional QRs, the persistent current is given by
$j = e \hbar / (2 m^* i)\,( \Psi^* \nabla \Psi - \Psi \nabla \Psi^*) 
- (e^2/m^*) A |\Psi |^2 = e / (2\pi m^* R)\,\left[ \hbar m_z/R - e A \right]
= \hbar^2/(m^* \Phi_0 R^2) (m_z+\Phi/\Phi_0)$, where $e$ is the electron 
charge, $R$ the ring radius, $\Phi$ the magnetic flux and $\Phi_0$ the
unit of flux.
\bibitem{ow} In other words, the coupling between left and ring wavefunctions
is larger when the localized states have opposite angular momenta
(antiparallel field), and smaller when they have the same angular momenta
(parallel field). The larger $|\langle m_z \rangle_l|$ and 
$|\langle m_z \rangle_l|_r$, the more so.
\end{thebibliography}
\end{document}